# How Electrons Become Mobile in a Colossal Dielectric – $Fe_2TiO_5$

M. L. McLanahan and A. P. Ramirez

*Physics Department, University of California Santa Cruz, Santa Cruz, CA, 95064*

ABSTRACT

We measure the colossal permittivity in single crystal $Fe_2TiO_5$ using broadband spectroscopy in the frequency range 20 Hz - 1 MHz. The relaxation response is analyzed using a Debye-like model with Arrhenius activation in two different ways and yields an energy barrier of 286.1 $\pm$ 2.8 meV. DC transport yields an activation energy of 288.8 $\pm$ 2.8 meV. These results strongly imply that the energy barrier for localized dipole motion and itinerant charge transport originate from the same atom-level forces. A further implication is that colossal dielectric behavior is a microscopic bulk phenomenon arising from a system on brink of metallicity.

**Introduction**

The conversion of electrons from localized to itinerant states is often discussed in terms of a metal-insulator transition (MIT). Systems on the insulating side display an increase in dielectric constant as the MIT is approached, driven either by a decrease in on-site potential, an increase in the tight-binding hopping parameter, or by a decrease in disorder. Such delocalization eventually gives way to hopping conductivity while still on the insulating side. On the metallic side of the transition, electrons are described by Bloch-like, delocalized wavefunctions which have lost the details of the insulating-side dynamics. Thus, to study the *dynamics* of delocalization one needs to remain on the insulating side and interrogate the transition from localized charge to mobile charge transport.

Important quantities for characterizing localized-to-mobile dynamics are *i*) the energy barrier governing charge motion within the primitive structure of an electric dipole and *ii*) the characteristic energy scale for charge hopping, immediately after breaking free of the primitive unit. One of the problems encountered in attempting to measure both of these barriers is material selection and sample control. The material to be studied should exhibit both dielectric response and hopping transport and, to minimize extrinsic effects, single crystals with well-characterized disorder are required

For this study, we chose $Fe_2TiO_5$ because it possesses a so-called "colossal" dielectric constant ($\varepsilon$) indicating proximity to a MIT. The dielectric constant is important for both microelectronics and sensing applications and, whereas conventional dielectric materials such as $SiO_2$, $HfO_2$, and $BaTiO_3$ exhibit room temperature $\varepsilon$ values of 3.9, 25, 1000 respectively, colossal $\varepsilon$ values e.g. in $CaCu_3Ti_4O_{12}$ (CCTO) [1] exhibit $\varepsilon \gtrsim 10^4$. Early work on $Fe_2TiO_5$ crystals also revealed a colossal dielectric response while possessing a band gap of ~ 2.2 eV, making it a candidate for photocatalytic water splitting [2-4]. Here we explore the frequency and temperature dependence of $\varepsilon$ needed to extract energy scales characterizing the dynamics of localized and itinerant charge.

Colossal dielectric response in materials like CCTO and $Fe_2TiO_5$ is accompanied by non-trivial frequency response evocative of relaxor ferroelectrics [5-7]. Like relaxors, $Fe_2TiO_5$ possesses intrinsic disorder due to random occupation of the two cation sites by $Fe^{3+}$ and $Ti^{4+}$. Thus, the Ti ion responsible for dipole moments occupies the two cation sites at 33% density. The magnitude of the dielectric response in some colossal dielectrics, however, has been attributed to interface effects such as electrode polarization or Maxwell-Wagner-type polarization [8-10] as in CCTO [11]. Determining the origin of the colossal behavior is necessary for device implementation and often requires the use of multiple formalisms for its full analysis. In particular, we focus here on the dielectric and impedance frameworks to characterize the response in $Fe_2TiO_5$. We find that the dielectric response is composed of three contributions: extrinsic interface, bulk relaxation, and conduction. This shows up as two easily resolvable peaks in the impedance spectra, the lower frequency one originating from overlapping extrinsic and bulk contributions, and the higher frequency peak related to the conductivity. Furthermore, our analysis supports the existence of *intrinsic dielectric response* of $Fe_2TiO_5$ which in indeed colossal, even after accounting for the extrinsic contact interface contribution.

$Fe_2TiO_5$ crystallizes in the $A_2BO_5$ pseudobrookite structure with space group Cmcm. This structure shares some similarity to CCTO, notably in its $TiO_6$ octahedra and multiple $A$-site cations, but whereas CCTO has an ordered arrangement of $Ca^{2+}$ and $Cu^{2+}$ cations, $Fe_2TiO_5$ possesses intrinsic disorder as already mentioned, so that the electric dipoles associated with $Ti^{4+}$ are randomly distributed. To better understand the local charge dynamics, we studied the broadband frequency-dependent dielectric response and explored its interplay with dc charge transport.

We measured the capacitance, $C$, and loss tangent (dissipation factor), $\tan(\delta)$, of $Fe_2TiO_5$ single crystals in the frequency range of 20 Hz – 1 MHz via an Agilent 4284A LCR meter and in the temperature range of 5 K – 325 K using a Quantum Design PPMS. To prepare $Fe_2TiO_5$ samples (obtained from the UCSC crystal archive [12]) for dielectric measurements, crystals were mechanically polished into rectangular plates, and Au (45 nm)/Cr (5 nm) electrodes were thermally evaporated onto the two faces, forming a parallel plate capacitor. Electrodes were deposited either parallel or perpendicular to the $c$-axis, between which we find a small difference in dielectric response (see supplemental material [13]). Several samples were measured to assure reproducibility, with dielectric data presented probing the response in the $ab$-plane (dimensions $A = 1.50$ mm $\times$ 2.00 mm and $d = 0.67$ mm, where $A$ is the area of the electrodes and $d$ the thickness of the crystal). Four-wire resistance measurements were made in a Quantum Design PPMS from 210 K – 350 K on a sample with dimensions 3.00 mm $\times$ 1.00 mm $\times$ 0.50 mm.

**Results**

The complex dielectric constant, $\tilde{\varepsilon}(\omega, T) = \varepsilon' + i\varepsilon''$, was determined from $C$ and $\tan(\delta)$ using: $\varepsilon' = C/C_0$ and $\varepsilon'' = \varepsilon' \tan(\delta)$, where $C_0 = \epsilon_0 A/d$ and $\epsilon_0$ is the permittivity of free space.

We plot the temperature and frequency dependence of $\varepsilon'$, $\varepsilon''$, and $\tan(\delta)$ in Fig. 1. A colossal dielectric constant of $\varepsilon' > 10^4$ is observed for $f \leq 10$ kHz at room temperature. On cooling, $\varepsilon'$ decreases to an asymptotic value of ~100. The diminishing response in $\varepsilon'$ is accompanied with a peak in $\tan(\delta)$ and is consistent with previous results for polycrystalline samples [14]. The temperature range in which relaxation behavior is seen, $170 \text{ K} < T < 300 \text{ K}$, is determined indirectly by the low and high-frequency limits respectively of our LCR meter. The drop in signal observed in $\varepsilon'$ shifts to lower frequencies upon cooling, implying a thermally activated relaxation process with relaxation time, $\tau(T)$. This behavior is not mirrored, however, by corresponding peaks in $\varepsilon''$ as would be expected for Debye-type relaxation i.e.,

$$\tilde{\varepsilon}(\omega) = \varepsilon_\infty + \frac{\Delta\varepsilon}{1 - i\omega\tau_D}, \tag{1}$$

where $\tau_D$ is the Debye relaxation time and $\Delta\varepsilon$ the dielectric relaxation strength. For a single Debye relaxation, $\Delta\varepsilon = \varepsilon_s - \varepsilon_\infty$, for which $\varepsilon_s$ and $\varepsilon_\infty$ are the low and high frequency dielectric constants respectively. Instead of a peak, as Eq. 1 anticipates, a shoulder-like bend can be seen in $\varepsilon''$ as a function of frequency at elevated temperatures, suggesting that the loss peaks are obscured by the presence of an additional contribution to the dielectric function.

This capacitance behavior already illustrates the complexity of extracting materials parameters from relaxation data [15], a complexity that results from a lack of microscopic knowledge of the spatial coordination of the physical structures within the device that map onto circuit elements, here capacitive and resistive. These microscopic structures must be lumped together in the circuit representation and then configured in either series of parallel configurations that most appropriately render the physical system. In particular, as colossal dielectric behavior has previously been ascribed to interfacial polarization, e.g. electrode and Maxwell-Wagner polarization effects, it is important to perform the analysis in a manner that doesn't obscure the intrinsic response of the material, but rather disentangles different response components.

As a first step in determining the appropriate circuit for modeling the intrinsic physics, we show that the shoulders in $\varepsilon''(\omega)$ relaxation data correspond to peaks in $\tan(\delta)$, as shown in Fig. 1d. The temperature dependence of $\tau = 1/\omega_\delta$ displays Arrhenius activated behavior (see Fig.1d inset), i.e. $\tau = \tau_0 \exp(E_A/k_B T)$, where $\tau_0$ is the characteristic relaxation time, $E_A$ the activation energy, and $k_B$ the Boltzmann constant, and the fits yield $\tau_{0,\delta} = (26.6 \pm 2.6)$ ps and $E_{A,\delta} = (291.4 \pm 1.6)$ meV, comparable to values previously reported for polycrystalline $Fe_2TiO_5$ samples ($\tau_0 \approx 50$ ps, $E_A = 140$ meV) [14]. While $\tan(\delta)$ analysis provides a coarse characterization of the thermally activated process, it is important to ask if these features are the result of a composite process.

To gain insight into the possible hidden contributions to the overall dielectric response, we fit $\varepsilon'$ and $\varepsilon''$ isotherms for T = 300 K to 245 K using a Cole-Cole model [16], an empirical extension of the Debye model (Eq. 1) that accounts for a distribution of relaxation times centered around $\tau$ instead of a single characteristic relaxation time, combined with a dc conduction term. For T < 245 K, the conductivity shifts to a power law at high frequencies, which is not captured in this simple model. While such a single-relaxation-process model describes the overall trend of the data, we find the fit is improved upon the addition of a second Cole-Cole term i.e.,

$$\tilde{\varepsilon} = \varepsilon_\infty - \frac{i\sigma_0}{\epsilon_0 \omega} + \frac{\Delta\varepsilon_1}{1+(i\omega\tau_1)^{\alpha_1}} + \frac{\Delta\varepsilon_2}{1+(i\omega\tau_2)^{\alpha_2}}, \qquad (2)$$

where $\sigma_0$, and $\alpha$ correspond to the conductivity at $\omega = 0$, and the Cole-Cole exponent, respectively. The improvement in fit is statistically significant according to both the partial F-test for nested models, in which the F-statistic compares the improvement in fit from additional parameters with the mean square error, as well as the Akaike information criterion (AIC), which compares model fits by penalizing models with more parameters [17] (see supplemental [13]). We show a representative fit for T= 300 K with individual dielectric contributions in Fig. 2a, with fit parameters and other isotherms presented in the supplemental. Because the data are best described by two relaxation terms and are measured on single crystals, we interpret the higher-frequency relaxation as a bulk response and the lower-frequency relaxation as a possible electrode effect. Next, we turn to the complex impedance, a formalism better suited to assessing electrode/interface contributions.

The complex impedance, $\tilde{Z}$, a quantity often used to analyze dielectric relaxation, is related to $\tilde{\varepsilon}$ via the relation

$$\tilde{Z} = Z' + iZ'' = (i\omega C_0 \tilde{\varepsilon})^{-1}, \qquad (3)$$

Whereas $\tilde{\varepsilon}$ emphasizes parallel processes, $\tilde{Z}$ emphasizes processes better suited to a series configuration, an appropriate rendition of a physical situation involving electrodes [15]. Specifically, electrode contributions, which we denote as $\tilde{Z}_{\text{Elec}}$, are often represented using a constant phase element (CPE) in a circuit block which is added in series to the bulk impedance [18, 19]. A CPE approximates nonideal capacitive behavior and has an impedance $\tilde{Z}_{\text{CPE}} =$ $1/Q(i\omega)^n$, where $Q$ is an effective capacitance, and $n$ interpolates between an ideal capacitor ($n = 1$) and an ideal resistor ($n = 0$). In Fig. 2b we show a Nyquist plot ($Z''$ vs $Z'$), where each point is a different frequency, of isotherms in our earlier fitting window. The presence of two semicircles demonstrates at least two relaxation processes, also represented as two peaks in $Z''$ vs $\omega$ (see supplemental for full $Z'$ and $Z''$ spectra [13]). Such a spectrum is often interpreted by assigning the higher frequency process to a bulk response and the lower frequency one to an extrinsic/barrier-layer contribution. Following the same process used for $\tilde{\varepsilon}$, we fit our data in the impedance formalism to $\tilde{Z} = \tilde{Z}_{\text{Elec}} + \tilde{Z}_{\text{Bulk}}$. Here, $\tilde{Z}_{\text{Elec}}$ is modeled by an $R\|\text{CPE}$ circuit element, i.e. a resistor of resistance $R$ in parallel with a CPE, a form commonly used to describe interface effects that appear as depressed semicircles in Nyquist plots [19], whereas $\tilde{Z}_{\text{Bulk}}$ is taken as the impedance of Eq. 2 without the second relaxation term. Explicitly, the impedances are

$$\tilde{Z}_{\text{Elec}} = \left[\frac{1}{R} + Q(i\omega)^n\right]^{-1}, \tag{4}$$

$$\tilde{Z}_{\text{Bulk}} = \left[(i\omega C_0)\left(\varepsilon_\infty - \frac{i\sigma_0}{\epsilon_0 \omega} + \frac{\Delta\varepsilon}{1+(i\omega\tau_D)^\alpha}\right)\right]^{-1}. \tag{5}$$

Fit curves are shown in Fig. 2b and in our case, we find the high-frequency (left-most) semicircle is associated with the interplay between $\sigma_0$ and $\varepsilon_\infty$, whereas the lower-frequency one reflects the overlap of the bulk and electrode relaxation responses. This can be seen explicitly in Fig 2c where we plot a representative fit for 300 K, showing both $\tilde{Z}_{\text{Elec}}$ and $\tilde{Z}_{\text{Bulk}}$ contributions. Fig. 2d presents Arrhenius plots of the fit relaxation times $\tau_D$ and $\tau_{\text{Elec}} = (RQ)^{1/n}$ yielding characteristic relaxation times $\tau_{0,D} = (786 \pm 175)$ ps and $\tau_{0,\text{Elec}} = (49.0 \pm 4.7)$ ns with activation energies $E_{A,D} = (280.7 \pm 5.3)$ meV and $E_{A,\text{Elec}} = (234.6 \pm 2.3)$ meV. Most notable from this analysis we find $\Delta\varepsilon > 10^4$ for all isotherms fit (see supplemental for fit parameters [13]), confirming a colossal dielectric response even after accounting for electrode contributions. The similar magnitudes of $E_{A,D}$ and $E_{A,\text{Elec}}$ are responsible for the intertwining of bulk and electrode responses but can be seen to arise from the same underlying process, discussed below.

**Conductivity Analysis**

Next, we rationalize the above relaxation analysis with both low- and high-frequency limiting behavior. The dc resistivity from 350 K – 210 K is presented in Fig. 3b alongside the resistivity from impedance fits, with agreement between the two within error bars. While $Fe_2TiO_5$'s intrinsic disorder would suggest variable range hopping, the limited temperature range allows the extraction of an energy barrier by assuming semiconductor temperature dependence [20], i.e.

$$\rho_{dc} = \rho_0 \exp(E_{A,\rho}/k_B T), \tag{6}$$

fits the dc data well with parameters $\rho_0 = (0.21 \pm 0.03)$ Ω-cm and $E_{A,\rho} = (288.8 \pm 2.8)$ meV. We note this activation energy is statistically identical to that found for $\tan(\delta)$ and near $E_{A,D}$ from the impedance fits, *implying that mobile charge carriers as well as the localized, Debye-relaxing charges, are governed by the same energy barrier*.

Finally, we address the high-frequency region, where Jonscher's universal dielectric response formalism allows extraction of an energy barrier for charge transport. To compare the dc-transport to the ac-dielectric measurements, we consider the ac-conductivity, which is related to the dielectric loss by

$$\sigma_{ac} = \omega \epsilon_0 \varepsilon''. \tag{7}$$

We plot $\sigma_{ac}$ in Fig. 3a and from these isotherms, we observe four distinct regions: *i*) a low-frequency plateau at high temperatures, *ii*) a dispersive region at frequencies above the low-frequency plateau and leading to *iii*) a second $\sigma_{ac}$ plateau, and *iv*) a high-frequency dispersive region at low temperatures. Regions *i* - *iii* can be captured using the impedance model discussed above but is inadequate to capture region *iv*. The high-frequency dispersive region (*iv*) is attributed to the "universal" dielectric response proposed by Jonscher [21], i.e. a power law which describes $\sigma_{ac}(\omega)$ in dielectric materials:

$$\sigma_{ac} = \sigma_{plat} + b\omega^s, \tag{8}$$

where $b$ is a constant, and $s$ is an exponent ranging from 0 to 1 whose temperature dependence is related to the dominant method of charge transport. Here, we use $\sigma_{plat}$, conventionally written as the dc conductivity, to denote the conductivity plateau preceding region *iv*. Typical conduction mechanisms include quantum mechanical tunneling (QMT), non-overlapping small polaron tunneling (NSPT), and correlated barrier hopping (CBH) [22, 23]. QMT is characterized by a temperature independent $s \approx 0.8$ and is prevalent at low temperatures. In contrast, $s$ increases with temperature for NSPT, while a decreasing $s$ with increasing temperature is observed in CBH. To identify the conduction mechanism in our system, we perform a linear fit of $\sigma_{ac}(\omega) - \sigma_{plat}$ versus $\omega$ on a log-log plot, where here $\sigma_{plat}$ is estimated from region *iii* (see supplemental for fitting details [13]). The slope obtained from this fit directly yields $s$.

The results of the fits are highlighted in Fig. 4, where we have scaled the data by Jonscher's equation such that we plot $(\sigma_{ac} - \sigma_{plat})/\sigma_{ac}$ vs. $(b/\sigma_{plat})\omega^s$ for 175 K $\leq T \leq$ 215 K, i.e. the temperature region where both region *iv* is observable and we can effectively estimate $\sigma_{plat}$. We find for $(b/\sigma_{plat})\omega^s \gtrsim 0.8$, the isotherms coalesce onto a master line, indicating the onset of the 'universal' dielectric response regime. In the inset of Fig. 5 we plot $s(T)$, which increases with decreasing temperature. This temperature dependence is a signature of CBH, a conduction mechanism where charge carriers (e.g. electrons or polarons) hop between localized states over a potential barrier, with a barrier height dependent on hopping distance due to the overlapping Coulomb potential wells from neighboring sites [24, 25], which is consistent with our earlier comments on transport in this disordered system.

| Measurement | Analysis | Energy Barrier (meV) |
|---|---|---|
| $\varepsilon(T,\omega), \tan(\delta)$ | Arrhenius | $291.4 \pm 1.6$ |
| $\varepsilon(T,\omega), Z_{Bulk}$ Fit | Arrhenius | $280.7 \pm 5.3$ |
| $\rho_{dc}$ | activation energy | $288.8 \pm 2.8$ |

Table 1. Activation energies extracted from dielectric and resistivity measurements.

**Discussion**

In summary, we measured the dielectric response as a function of frequency and temperature as well as the dc-conductivity temperature dependence of Fe$_2$TiO$_5$. We considered different circuit analogues and showed that the low-frequency relaxation process often ascribed solely to extrinsic mechanisms is better described by two processes, a likely electrode contribution in addition to an intrinsic contribution of comparable magnitude. The equivalence of the energy barriers for each process suggests that a single process, namely the dipole moment reversal barrier at the unit cell level, underlies both dielectric relaxation and charge transport. Thus, the analysis presented here implies that mobile charge carriers *should* accumulate at the interface of electrodes for large enough conductivity thus giving rise to a low frequency process accompanying the bulk response. Future investigation of the relaxation activation energy (~286 meV) may be of interest, as it likely originates from lattice deformation. Our results paint a different picture, though, from a previous explanation of the origin of colossal dielectric behavior in CCTO which attributed the large dielectric response solely to a Maxwell-Wagner effect at either the electrodes or at ill-defined internal barrier layers [11]. Thus, for Fe$_2$TiO$_5$, the results presented here show that the high dielectric constant can be driven by intrinsic large dipole moments created by the proximity to a MIT. While a bulk dielectric response may be inseparable from the electrode effect in practical applications, its origin in an intrinsic materials property suggests controllable engineering approaches towards device integration.

Acknowledgements –  We acknowledge useful discussions with T. Siegrist. This work was supported by the U. S. Department of Energy Office of Basic Energy Sciences, division of Condensed Matter Physics grant no. DE-SC0017862.

FIGURES

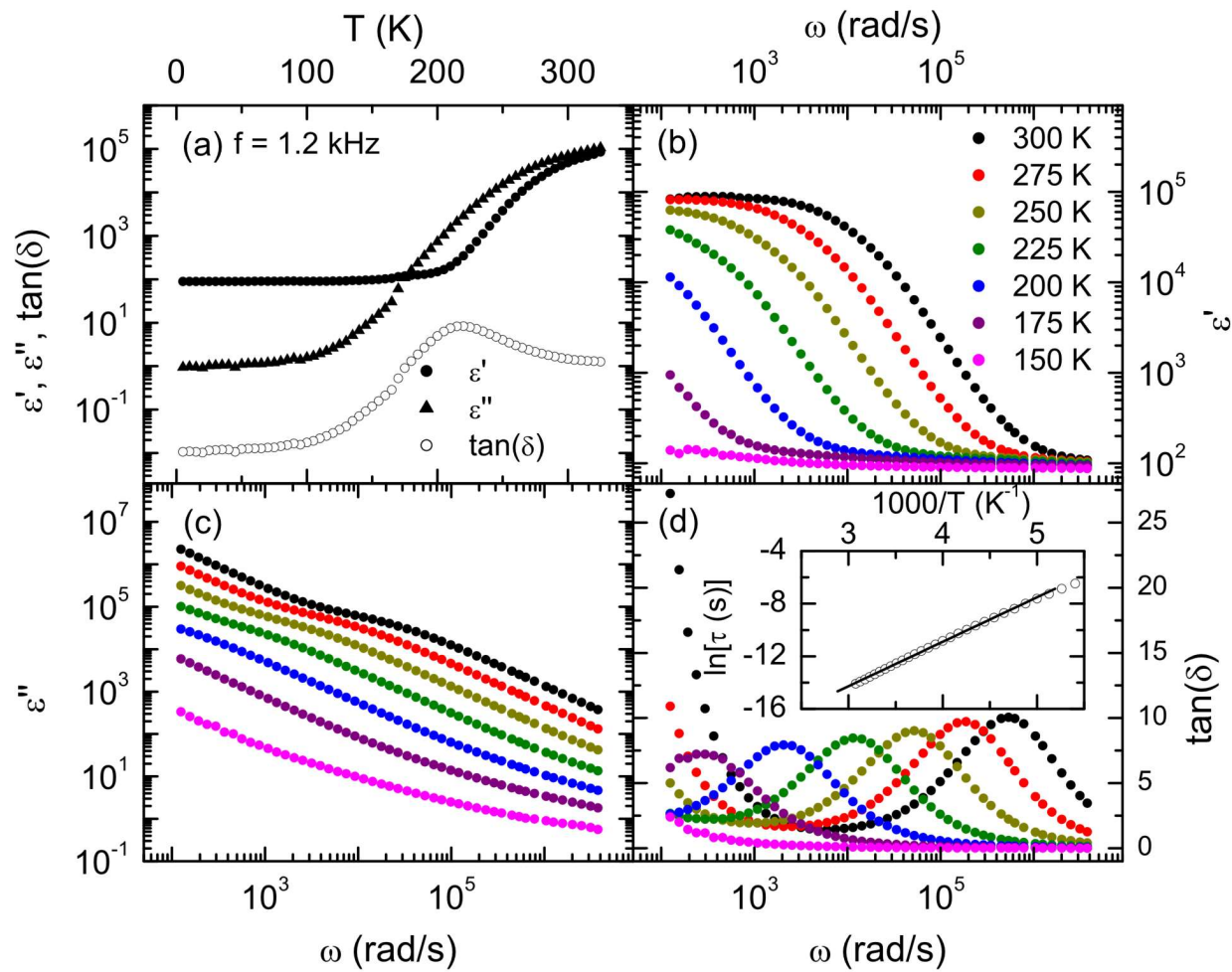

Figure 1. (a) temperature dependence of $\varepsilon'$, $\varepsilon''$, and $\tan(\delta)$ for single crystal $Fe_2TiO_5$ at $f = 1.2$ kHz with drive field perpendicular to the crystallographic $c$-axis. Large $\varepsilon'$ at room temperature decreases with decreasing temperature accompanied by a loss peak in $\tan(\delta)$. Frequency dependence of (b) $\varepsilon'$, (c) $\varepsilon''$, and (d) $\tan(\delta)$ reveal dispersive behavior moves to lower frequencies upon cooling. Inset: Arrhenius plot of the inverse of the $\tan(\delta)$ loss peak frequency, $\tau$, with fit lines.

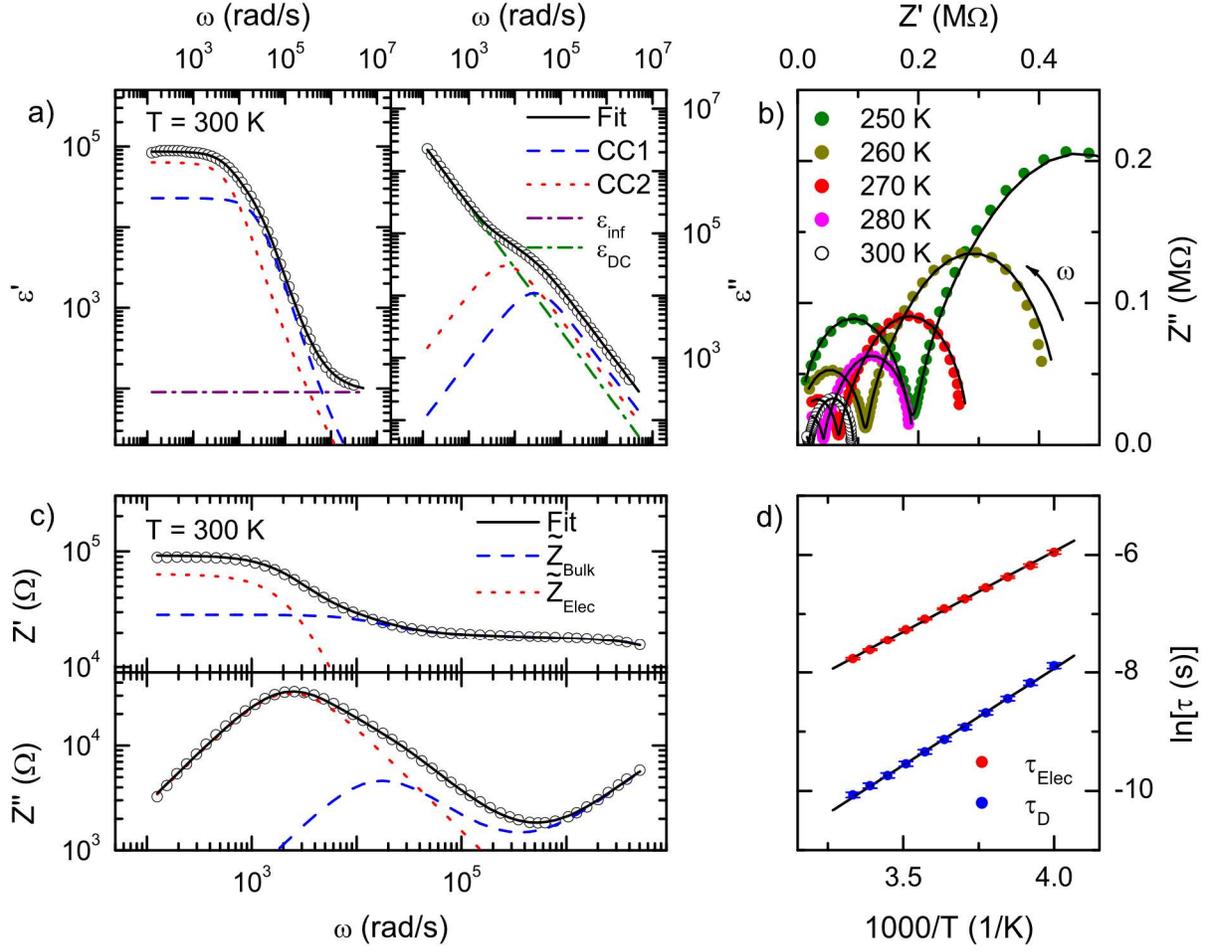

Figure 2. (a) real and imaginary parts of $\tilde{\varepsilon}$ measured at 300 K. Black solid lines correspond to fit curve from a 2 Cole-Cole + dc dielectric model, while the other curves are the individual contributions. (b) Nyquist plot where each point in an isotherm is a different frequency, black lines are from impedance fits. (c) real and imaginary parts of $\tilde{Z}$ measured at 300 K. Black solid lines correspond to fit curve, while the other curves are the individual contributions from the Cole-Cole term and electrode term (R||CPE circuit element). (d) Arrhenius fit of the relaxation times from impedance fits corresponding to $\tilde{Z}_{Elec}$ (red) and $\tilde{Z}_{Bulk}$ (blue).

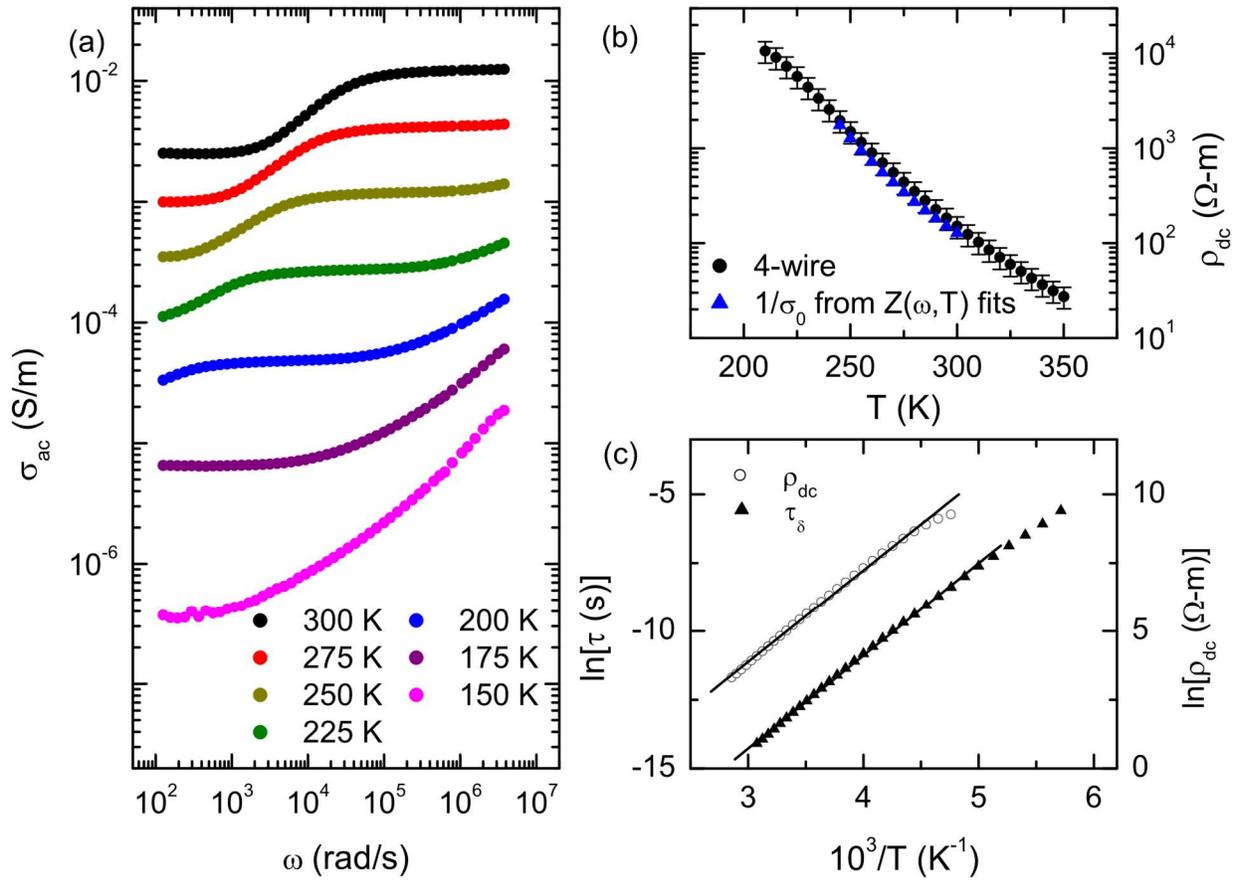

Figure 3 (a) $Fe_2TiO_5$ ac conductivity plotted as a function of frequency for the $ab$-plane direction. For $T > 250$ K there are two observable plateaus. At high frequencies and lower temperatures $\sigma_{ac}(\omega)$ undergoes dispersion described by a Jonscher power law. (b) Resistivity of $Fe_2TiO_5$ measured using a 4-wire setup (black) with error bars from 0.1 mm uncertainty in crystal dimensions, and $1/\sigma_0$ from impedance fit showing agreement. (c) Arrhenius plot comparing the activation energy (slope) of $\rho_{DC}$ and $\tau_\delta$.

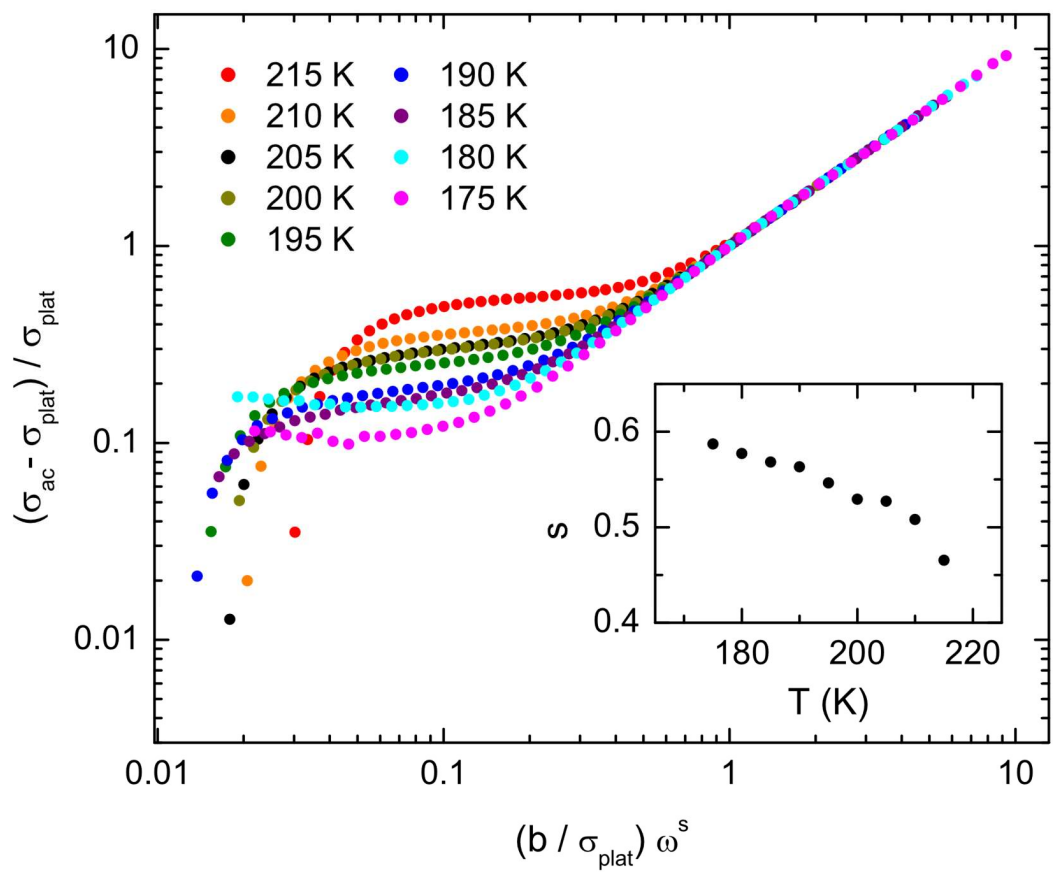

Figure 4. Ac conductivity scaled to Jonscher's power law. Inset: temperature dependence of the $\omega$ power, $s$. The inversely proportionate temperature dependence suggests dominant conduction method is correlated barrier hopping.

# Supplemental Note: How Electrons Become Mobile in Colossal Dielectric – Fe$_2$TiO$_5$

M. L. McLanahan and A. P. Ramirez

*Physics Department, University of California Santa Cruz, Santa Cruz, CA, 95064*

**Supplemental Note 1: Dielectric Response Crystal Orientation**

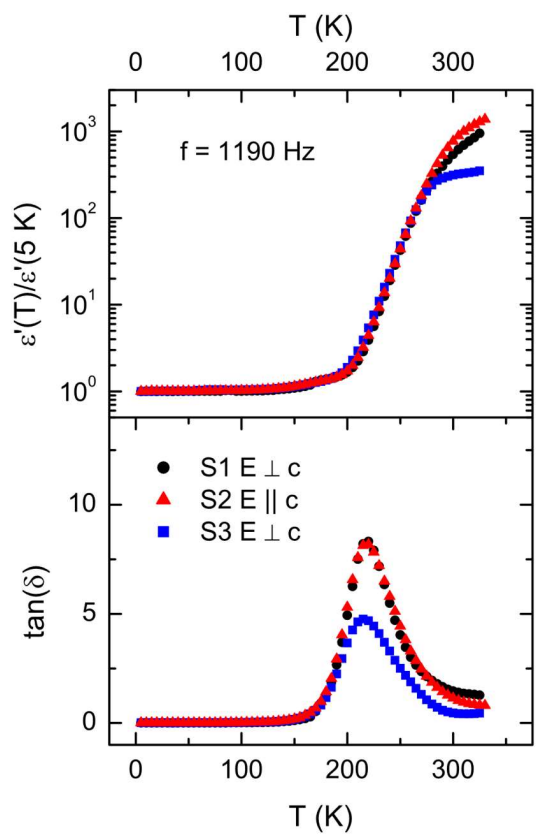

Figure S1. Scaled $\varepsilon'$ and $\tan(\delta)$ for three different samples, measured at $f = 1190$ Hz with excitation voltage parallel or perpendicular to crystal $c$-axis. Crystals were oriented based on magnetization data.

**Supplemental Note 2: Dielectric Fitting**

Dielectric spectra $\varepsilon'$ and $\varepsilon''$ were fit in log space using nonlinear regression to the two models described in the manuscript, i.e. Cole-Cole (CC) relaxation with a dc conduction contribution (model 1), and the same model with an additional CC term (model 2),

$$\tilde{\varepsilon}(\omega) = \varepsilon_\infty - \frac{i\sigma_0}{\omega\epsilon_0} + \sum_j \frac{\Delta\varepsilon_j}{1+(i\omega/\omega_{0,j})^{\alpha_j}}, \tag{S1}$$

where the last term represents the number of CC terms. Here, $\epsilon_0$ is the permittivity of free space, and the fit parameters $\varepsilon_\infty, \sigma_0, \Delta\varepsilon_j, \omega_{0,j}$, and $\alpha_j$ correspond to the high frequency dielectric constant, conductivity at $\omega = 0$, dielectric relaxation strength, relaxation frequency, and CC exponent for the $j^{th}$ CC relaxation term, respectively. Because model 1 is a nested version of model 2, we used the partial $F$-test to determine if the addition of another CC term statistically improved the fit. The partial $F$ statistic compares the reduction in residual error achieved by the more complex model with the reduction in error expected from additional parameters, and was calculated via

$$F = \frac{(RSS_1 - RSS_2)/(k_2 - k_1)}{RSS_2/(n - k_2)}, \tag{S2}$$

where $n$ is the number of points, $RSS_i$ the residual sum of squares, and $k_i$ the number of parameters for model $i$. The residuals for the two models are shown in Fig S1. For all isotherms fit, model 2 showed a large reduction in residual error relative to model 1, with an average $F = 433$. Under the null hypothesis that model 1 is sufficient, the corresponding $p$ value calculated from the $F$ statistic gives the probability of observing an improvement in fit at least this large by random chance, which we find $p < 10^{-40}$ for all isotherms, indicating that the improvement of model 2 over model 1 is statistically significant.

Furthermore, the two models were also compared using the small-sample corrected Akaike information criterion (AIC$_c$) which balances goodness of fit against model complexity, i.e. number of fit parameters, where a lower AIC$_c$ value indicates the better-supported model [1]. For least-squares fitting, AIC$_c$ is calculated by

$$\text{AIC}_c = n \ln\left(\frac{RSS}{n}\right) + 2k\left(\frac{n}{n-k-1}\right). \tag{S3}$$

Using the minimum-AIC$_c$ convention, in which the model with the lowest AIC$_c$ is assigned ΔAIC$_c$ = 0, model 2 had ΔAIC$_c$ = 0 and model 1 had an average ΔAIC$_c$ = 259. Because values of ΔAIC$_c$ > 10 are commonly taken to indicate essentially no support for a model relative to the best-supported model [1], model 2 is clearly favored over model 1.

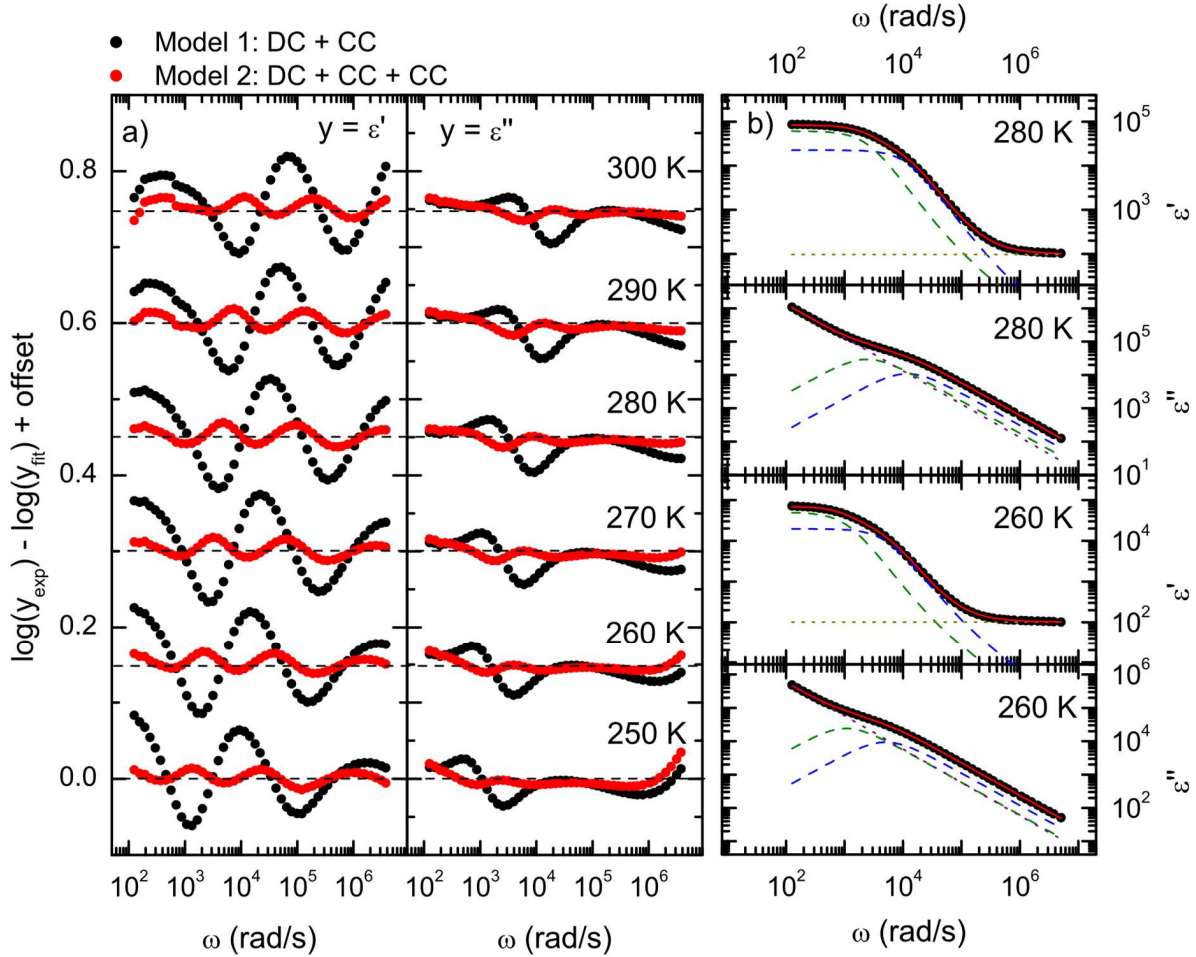

Figure S2. (a) Residual error from nonlinear regression of $\log(\varepsilon')$ and $\log(\varepsilon'')$ to model 1 (black) and model 2 (red). (b) $\varepsilon'$ and $\varepsilon''$ as a function of $\omega$ (black points) for $T = 280$ K and $T = 260$ K showing model 2 fit curves (red solid), with individual dielectric contributions: Cole-Cole term 1 (dashed green), Cole-Cole term 2 (dashed blue), $\varepsilon_\infty$ (dotted yellow), dc conduction term (dotted purple).

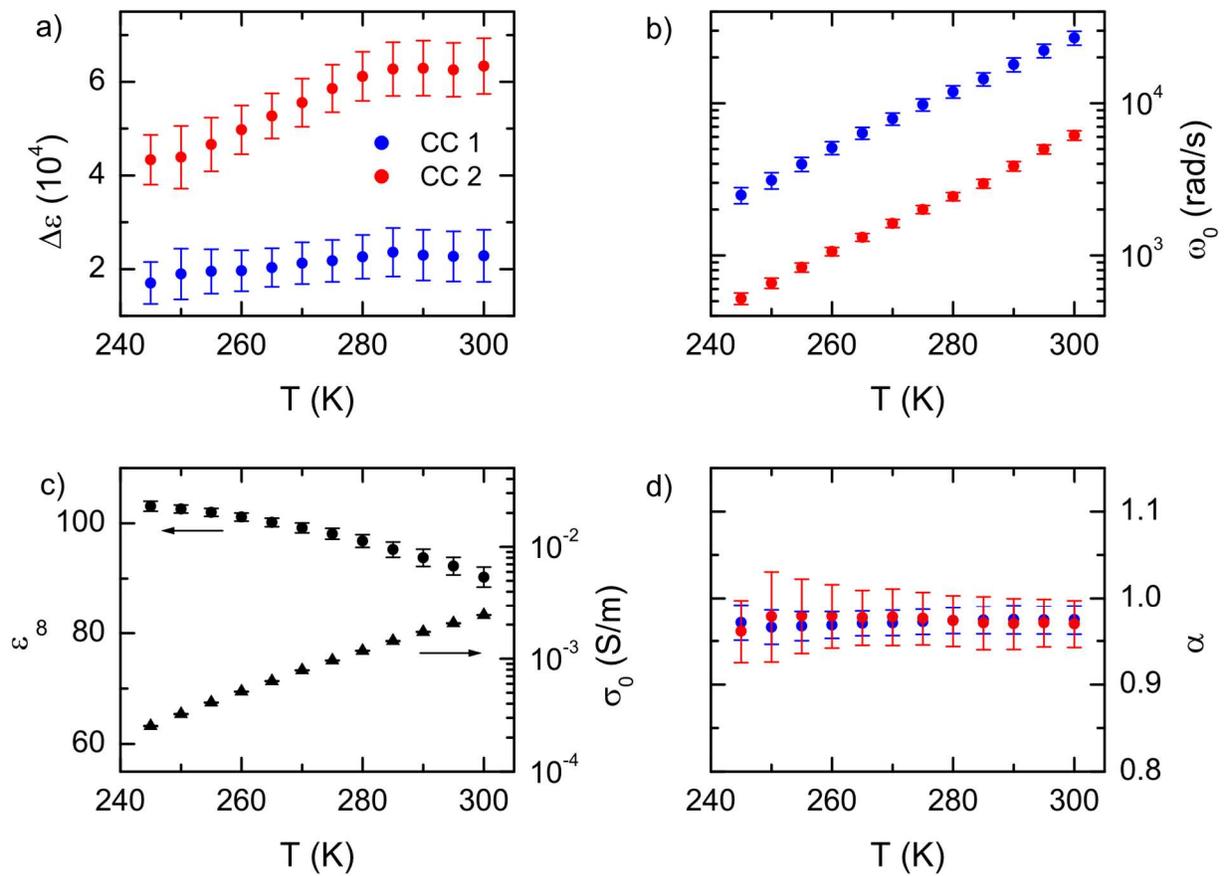

Figure S3. Fit parameters from fitting $\varepsilon'$ and $\varepsilon''$ to dielectric model 2, which includes 2 Cole-Cole relaxation terms with red and blue points corresponding to respective parameters from each term. Error bars are derived from standard errors from the nonlinear fit.

**Supplemental Note 3: Impedance Spectra**

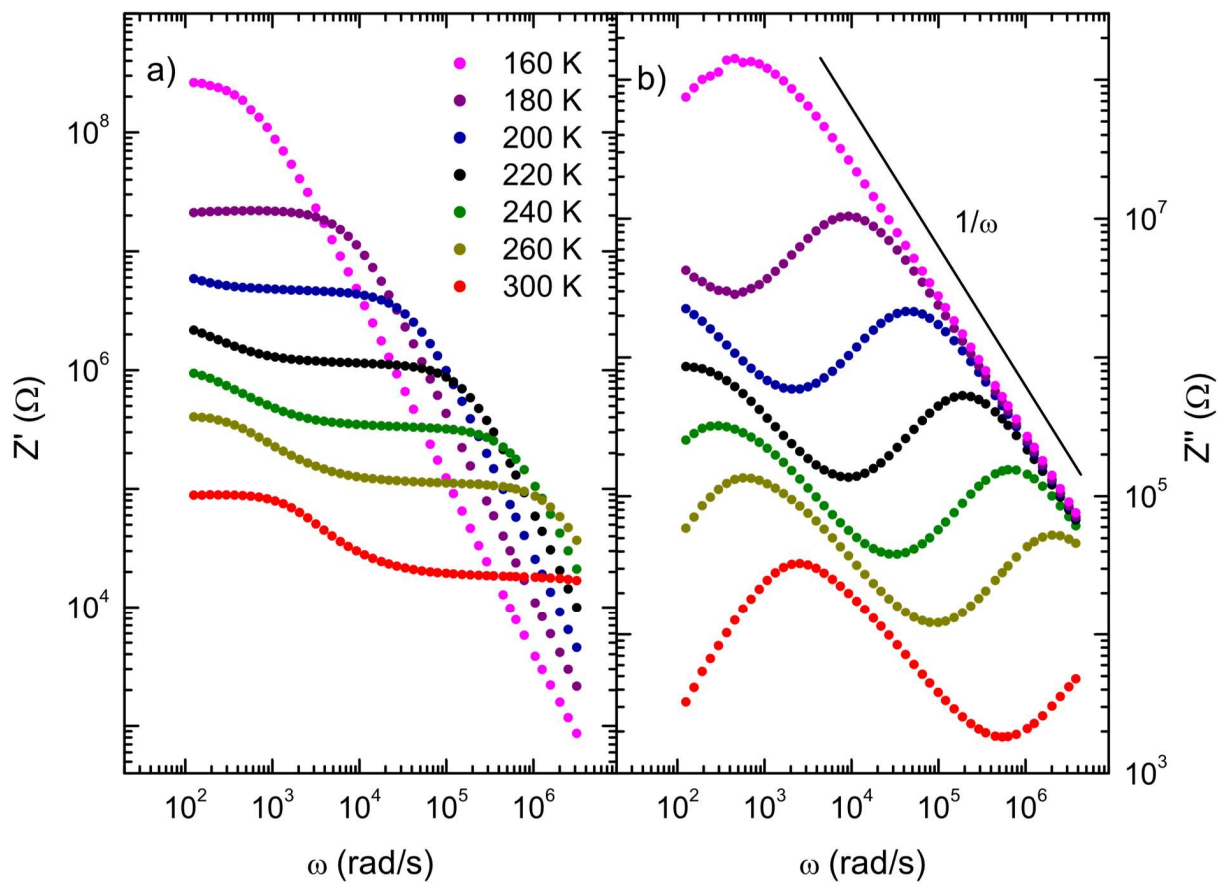

Figure S4. Real (a) and imaginary (b) impedance spectra, obtained from dielectric data. Two peaks in $Z''$ are clearly visible and shift to lower frequencies upon cooling.

**Supplemental Note 4: Impedance Fitting**

To better treat potential electrode/interface contributions, we worked in the impedance representation by transforming our dielectric data via $\tilde{Z} = (i\omega C_0 \tilde{\varepsilon})^{-1}$. As in the dielectric fitting, we fit $Z'$ and $Z''$ in log space using nonlinear regression to two impedance models, $\tilde{Z} = \tilde{Z}_{\text{Bulk}}$ (model 1) and $\tilde{Z} = \tilde{Z}_{\text{Bulk}} + \tilde{Z}_{\text{Elec}}$ (model 2) Here, $\tilde{Z}_{\text{Bulk}}$ is the impedance representation of dielectric model 1 from Supplemental Note 2, and $\tilde{Z}_{\text{Elec}}$ is the impedance of an R||CPE circuit element, i.e. a resistor in parallel with a constant phase element (CPE) used to model an electrode/interface contribution (see manuscript Eq. 4 and Eq. 5 for definitions). Impedance spectra were fit to both models for 300 K – 245 K, i.e. the temperature region where both $Z''$ peaks can be sufficiently resolved for both model fits. In Fig. S5 we plot fit curves of both models. Model 2 provides a better approximation of the data, which is also observed from the residual errors shown in Fig. S6. As model 1 is a nested version of model 2, we can again use the statistical tests from Supplemental Note 2 to quantitatively compare the fits of the two models. We find an average $F = 679$, $p < 10^{-39}$ for all fitted isotherms, $\Delta \text{AIC}_c = 0$ for model 2, and an average $\Delta \text{AIC}_c = 294$ for model 1, which strongly supports model 2 as a better description of our data, i.e. intrinsic bulk relaxation with dc conductivity and an additional interface contribution.

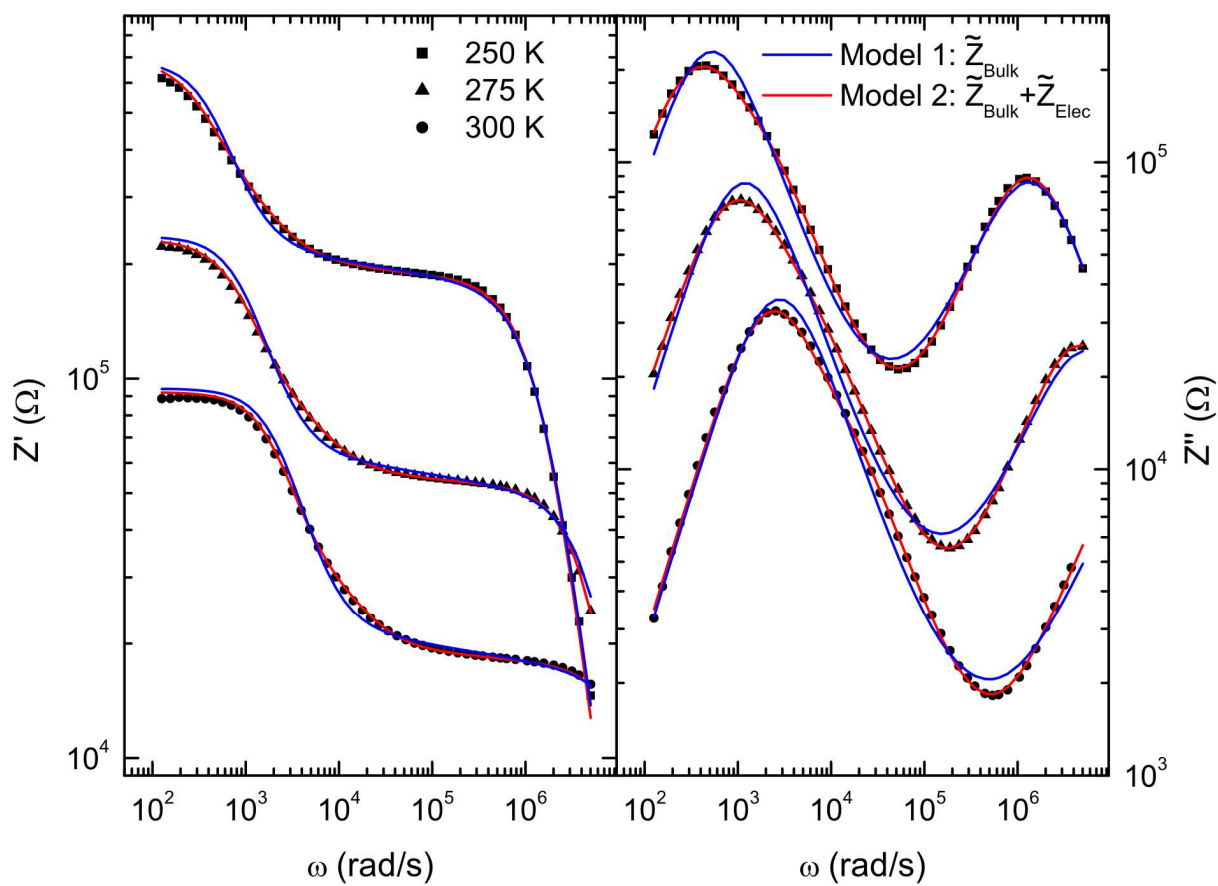

Figure S5. Impedance spectra from dielectric measurements showing impedance model 1 (blue) and model 2 (red) fit curves for different isotherms.

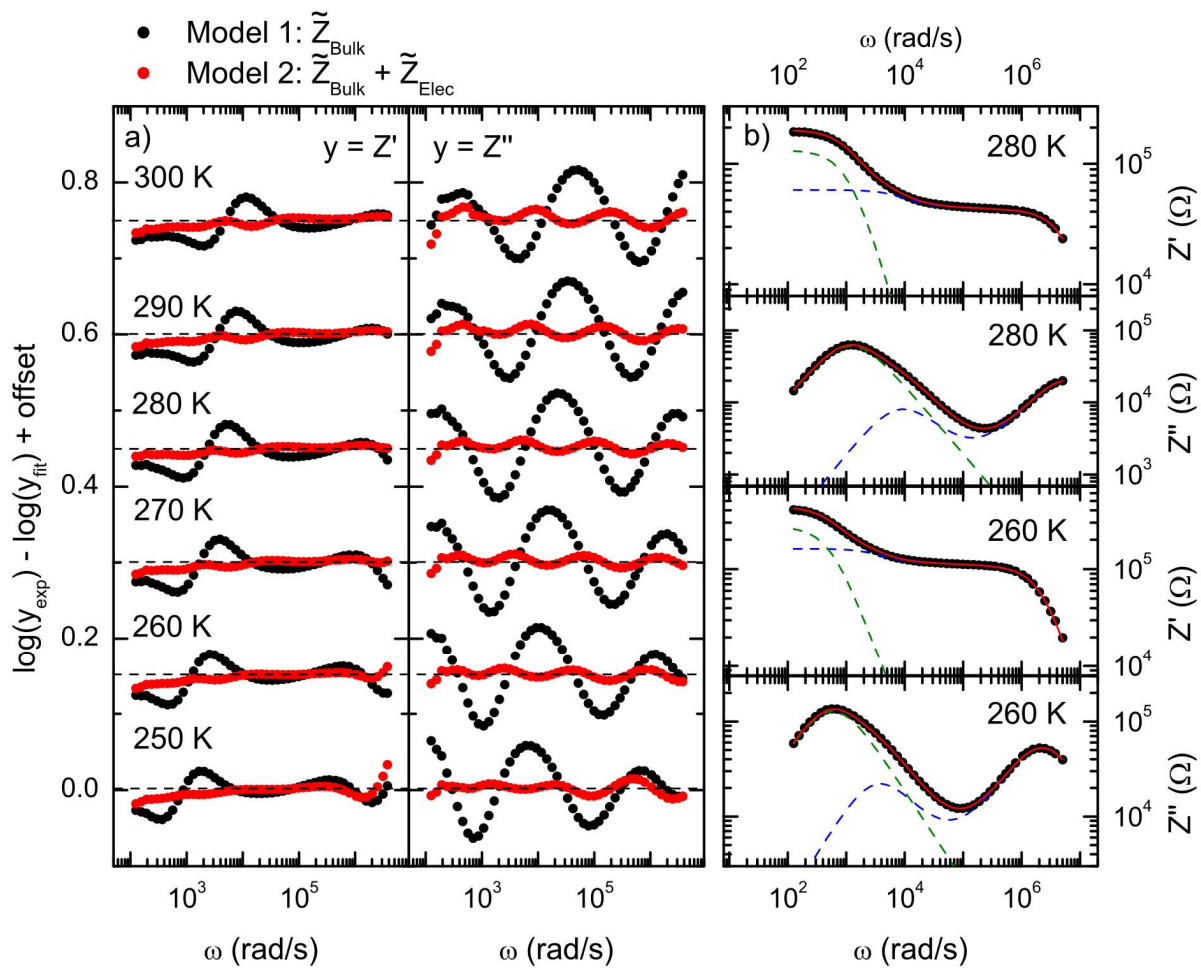

Figure S6. (a) residual error from nonlinear regression of $\log(Z')$ and $\log(Z'')$ to impedance models 1 (black) and 2 (red). (b) $Z'(\omega)$ and $Z''(\omega)$ (black points) for $T = 280$ K and $T = 260$ K with model 2 fit curves (red solid) and individual impedance components: electrodes (green dashed) and bulk (dashed blue) modeled as Cole-Cole relaxation with a dc conductivity term.

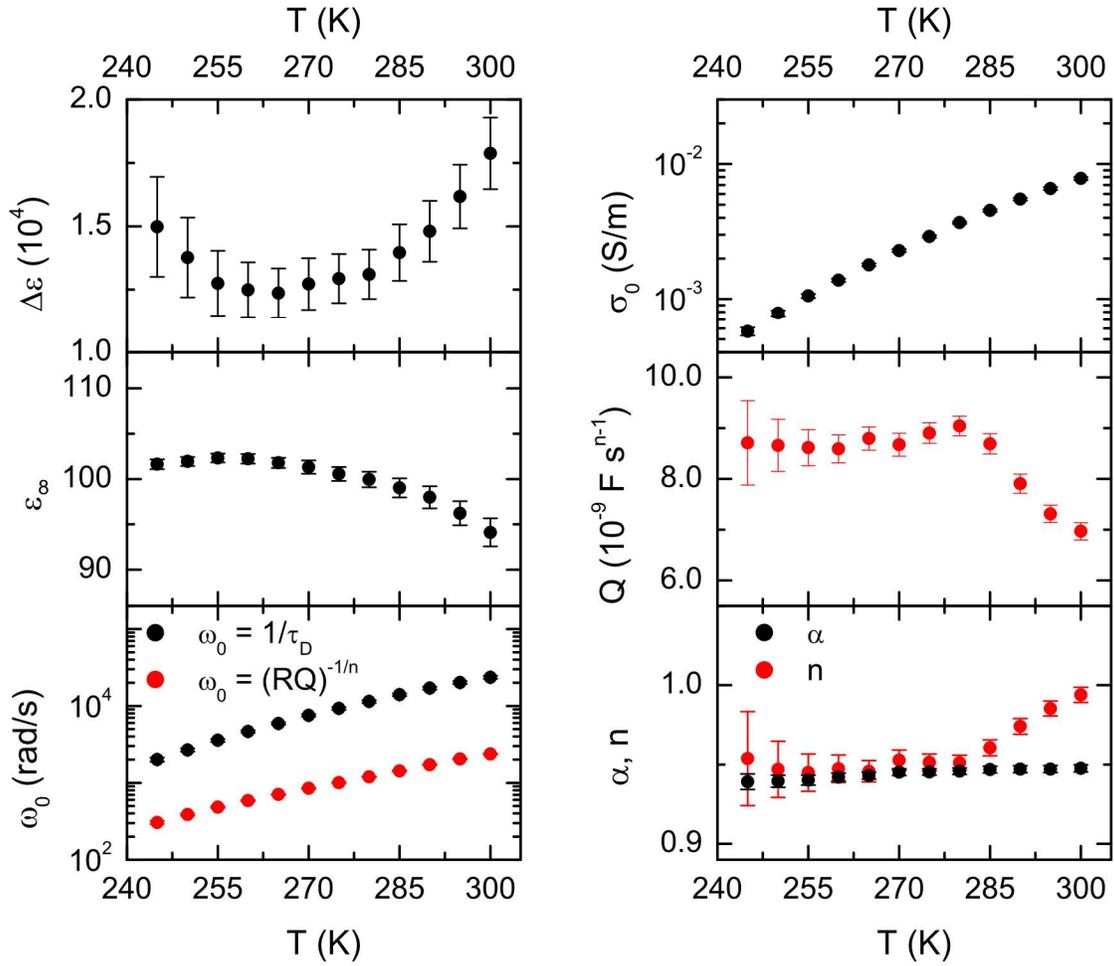

Figure S7. Fit parameters from fitting impedance spectra to impedance model 2. Red points correspond to parameters from $\tilde{Z}_{\text{Elec}}$ and black points to parameters from $\tilde{Z}_{\text{Bulk}}$. Error bars are derived from standard errors from the nonlinear fit. The error bars for $\omega_0$ and $\sigma_0$ are smaller than the plotted points.

**Supplemental Note 5: Conductivity Fitting**

When determining the temperature dependence of the Jonscher exponent $s$ by fitting ac-conductivity data to $\sigma_{ac} = \sigma_{plat} + b\omega^s$, care is required because $b$ and $s$ are not independently determined. Often the fitting frequency window is limited as well as temperature dependent, and variations in $b$ can partially compensate for variations in $s$ among the different isotherms. As a result, the apparent temperature dependence in $s$ may not reflect that of the material. Thus, a systematic and properly constrained approach is required, and it is necessary to check that all fitting parameters are well behaved. We outline our fitting procedure to determine $s$ below.

As described in the manuscript, we observe 4 different conductivity regions in $\sigma_{ac}(\omega)$: a low frequency plateau at elevated temperatures (*i*), a dispersive region (*ii*) leading to a second plateau at higher frequencies (*iii*), and at highest frequencies another dispersive region (*iv*). Because the lower-frequency process (regions *i* and *ii*) are not described by Eq. (8) but do influence region *iii*, we focus only on fitting region *iv*. In particular, we perform a linear fit on a log-log plot of $\sigma_{ac}(\omega) - \sigma_{plat}$ versus $\omega$, for which the slope directly yields $s$. To define the fit window systematically, we restrict the analysis to frequencies at least one decade above the $Z''$ higher-frequency peak (i.e. $\omega \geq 10\,\omega_Z$) for each isotherm, noting that the onset of the dispersion approximately coincides with $\omega_Z$ (see Fig. S8). Additionally, as the preceding plateau (region *iii*) contains contributions from both the dc-conductivity and lower frequency process, the value $\sigma_{plat}$ was treated as a constrained fit parameter and allowed to vary within the interval $[0.5, 2] \times \sigma_{plat}^{est}$, where our initial estimate is defined by the $\tan(\delta)$ peak frequency, i.e. $\sigma_{plat}^{est} = \sigma_{ac}(\omega_\delta)$. The optimal $\sigma_{plat}$ value was chosen that minimizes the reduced $\chi^2$ of the $\sigma_{ac}(\omega) - \sigma_{plat}$ versus $\omega$ linear fit, for which the corresponding slope was taken as $s$.

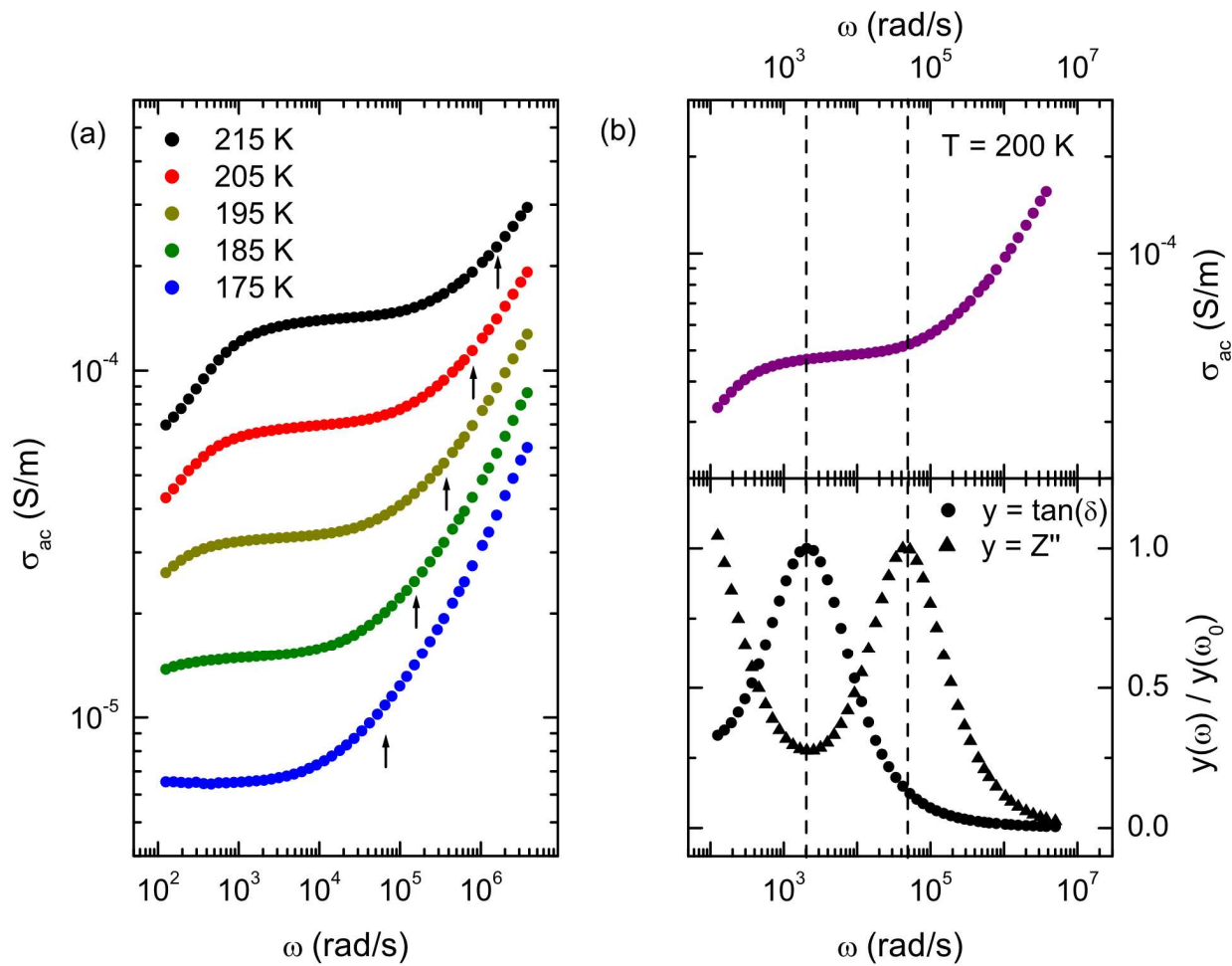

Figure S8. (a) ac conductivity ($\sigma_{ac}$) isotherms with arrows indicating the minimum frequencies considered in the Jonscher fits. (b) as an example of the general trend observed across isotherms, $\sigma_{ac}$ is shown alongside $\tan(\delta)$ and $Z''$, each normalized by peak frequency, demonstrating how these peaks coincide with $\sigma_{ac}$.